\documentclass[twocolumn,showpacs,preprintnumbers,amsmath,amssymb]{revtex4}
\usepackage{graphicx}
\usepackage{color}

\newcommand{\om}{\ensuremath{\omega}}
\newcommand{\omn}{\ensuremath{\omega_{n}}}
\newcommand{\im}{{\cal I}m\,}
\newcommand{\re}{{\cal R}e\,}
\newcommand{\DD}{\ensuremath{\delta}}
\newcommand{\cc}[1]{\ensuremath{C_{#1}}}
\newcommand{\ccp}[1]{\ensuremath{C_{#1}^{\prime}}}

\newcommand{\eqqref}[1]{Eq.~(\ref{#1})}
\newcommand{\beq}{\begin{equation}}
\newcommand{\eeq}{\end{equation}}
\newcommand{\beqn}{\begin{eqnarray}}
\newcommand{\eeqn}{\end{eqnarray}}

\begin{document}
\title{Dispersive spectrum and orbital order of spinless p-band
  fermions in an optical lattice
}

\author{Xiancong Lu and E. Arrigoni} 

\affiliation{Institute of Theoretical and Computational Physics, Graz
  University of Technology, A-8010 Graz, Austria}

\begin{abstract}
  We study single-particle properties of a spinless p-band correlated
  fermionic gas in an optical lattice by means of a variational cluster
  approach (VCA). The single-particle spectral function is almost flat at
  half-filling and develops a strongly dispersive behavior at lower
  fillings. The competition between different orbital orderings is studied as
  a function of filling.  
  We observe that an ``antiferromagnetic'' orbital order develops at half-filling and 
  is destroyed by doping the system evolving into a 
  disordered orbital state.
  At low filling limit, we discuss the possibility of ``ferromagnetic'' orbital order 
  by complementing the VCA result with observations based on 
  a trial wave function.
  We also study the behavior
  of the momentum distribution for different values of the on-site interaction.
  Finally, we introduce an integration contour in the complex plane
  which allows to efficiently
  carry out  
  Matsubara-frequency sums.
\end{abstract}

\maketitle

\section{Introduction}

Ultracold atomic gases in optical lattices constitute a promising system to
simulate and investigate strongly correlated quantum phases as a function of
their model parameters, which can be controlled experimentally in a large
range \cite{BlochRMP08}.  This field of research has greatly expanded after
the pioneering realization of the Superfluid to Mott-insulator transition by
loading bosonic atoms to the lowest band of optical lattices
\cite{Greiner02,Jaksch98}. Recent progress in this field has been achieved
by loading the atoms in the first excited band, which
makes the study of orbital physics possible in these systems
\cite{MullerPRL07,IsacssonPRA05,WLiuPRA06}.  Orbital degrees of freedom play
an important role in many solid-state materials: Many interesting phenomena
such as metal-insulator transitions~\cite{im.fu.98},
superconductivity~\cite{ka.wa.08}, colossal
magnetoresistance~\cite{da.ho.01}, half metallicity~\cite{gr.mu.83,ch.ar.06},
etc, are rooted in the coupling of
orbital with the other degrees of freedom (spin, charge, and phonon). The
study of orbital physics in optical lattices, in a pure and tunable
environment, is believed to be of great help to understand the complicated orbital
issue of solid-state systems.

The basic physics of cold atoms in the first excited band can be captured by
a p-band Hubbard like Hamiltonian \cite{IsacssonPRA05,WLiuPRA06}. Many novel
phenomena and quantum phases have been predicted for the p-band bosons
\cite{IsacssonPRA05,WLiuPRA06,CWuPRL06,KuklovPRL06,CWuPRL07,CXuPRB07}, e.g.,
quantum stripe order \cite{CWuPRL06}, Wigner crystallization \cite{CWuPRL07},
and bond algebraic liquid phase \cite{CXuPRB07}.  For the spin-1/2 p-band
fermions, an antiferromagnetic order was found at half filling in both
the strong
and weak interaction regimes \cite{KWuPRB08}, and a robust ferromagnetic
order was shown to exist for a large range of interaction and 
at band filling
lower than half-filling \cite{LWangPRA08}. Itinerant ferromagnetism was also
proposed in the honeycomb lattices in Ref.~\cite{SZhangarxiv08}.
Experimentally, the population of higher band was 
studied
by Browaeys
\textit{et al.} \cite{BrowaeysPRA05} and K\"{o}hl \textit{et al.}
\cite{Kohl05} for bosons and fermions respectively. Recent experiments
performed by M\"{u}ller \textit{et al.} were able to realize long lifetime p-band
orbital bosonic systems \cite{MullerPRL07}.

In particular, the orbital exchange physics in the Mott state of an orbital-only
model, which is realized by loading the single-component (spinless) fermions
into p-band optical lattices (see the Hamiltonian in Eq. (\ref{spbh})), has
been studied by Zhao \textit{et al.} \cite{EZhaoPRL08} and Wu \cite{CWuPRL08}
for various geometry lattices. In these works, a 
new orbital exchange mechanism was found, and
long-range orbital order was predicted. At the same time, a similar
orbital-only model was
proposed to describe the ferromagnetic plane in transition metal oxides with
$t_{2g}$ orbital degeneracy, such as $Sr_2VO_4$ and $K_2CuF_4$
\cite{da.wo.08}. The spectral properties of this model in the half-filled
case have been studied, and it was shown that a hole in a background of
antiferromagnetic orbital order does not localize but moves coherently due to
an effective three-site hopping term.

Motivated by these previous works, we  study numerically this spinless p-band
model on a square lattice 
with an emphasis on the excitation spectrum and orbital order away from
half-filling. The paper is organized as follows:
In Sec. \ref{vca}, we present the Hamiltonian of the model, and we  briefly
summarize the method used to approximately 
solve it, namely, the variational cluster approach. 
As a byproduct, in this work, we introduce and
adopt a more efficient method to 
carry out sums over Matsubara frequencies, which could also be applied
to other problems.
Details are given in
Appendix~\ref{freq}. In Sec. \ref{results}, we present the calculated results
including the single-particle spectrum, orbital orders, and momentum
distribution. Finally, we draw our conclusions in Sec. \ref{conc}.

\section{Hamiltonian and method}
\label{vca}

We consider an anisotropic 3D optical lattice with optical trapping
frequency $\omega_z \gg \omega_x=\omega_y$, so that the dynamics in the $z$
direction is essentially suppressed. 
Supposing that the lowest $s$ orbital of the optical
lattice is fully occupied by fermions, the other particles can only fill the
degenerate $p_x$ and $p_y$ orbitals \cite{EZhaoPRL08,CWuPRL08}. A fermionic
gas, which is polarized into a single hyperfine spin state by magnetic
field and loaded in such optical lattice, can be described by
the following 2D spinless p-band Hubbard Hamiltonian
\begin{eqnarray}\label{spbh}
H=\sum_{\mathbf{r}}\sum_{\alpha,\beta=x,y}t_{\alpha\beta}
          \Big( c_{\mathbf{r},\alpha}^\dag c_{\mathbf{r}+\mathbf{\hat{e}_\beta},\alpha} + H.c. \Big)
           +U\sum_{\mathbf{r}}n_{\mathbf{r},x}n_{\mathbf{r},y}
\end{eqnarray}
Here, $c_{\mathbf{r},\alpha}^\dag$ creates a fermionic atom in the $p_\alpha$ orbital at
position $\mathbf{r}$, $\mathbf{\hat{e}_\beta}$ is the unit vector of the $\beta$
direction 
(the lattice spacing is set equal to unity),
$t_{\alpha\beta}=t_{\parallel}\delta_{\alpha\beta}+t_{\perp}(1-\delta_{\alpha\beta})$
is the hopping amplitude of orbital $p_\alpha$ in direction $\beta$, and
$U$ is the on-site repulsive interaction between atoms in different orbitals.
The longitudinal hopping $t_{\parallel}$ is positive but the transverse
hopping $t_{\perp}$ is negative because of the odd parity of $p$ orbitals
\cite{EZhaoPRL08,CWuPRL08}.
In general, one has $|t_{\parallel}| \gg |t_{\perp}|$ for the strongly
anisotropic shape of $p$ orbital, therefore, we choose a typical small value
$t_{\perp}=-0.05t_{\parallel}$ in our calculation \cite{KWuPRB08,CWuPRL07}
and set $t_{\parallel}$ as the energy scale.  Obviously, in the limit
$t_{\perp}\to 0$ the hopping will be restricted to one dimension and the
number of particles in a given orbital ($p_\alpha$) will be conserved for
each chain oriented along $\alpha$.
There is no s-wave scattering for atoms in 
a single hyperfine spin state
because of the Pauli exclusion principle. Therefore, the interaction $U$ between
atoms mainly comes from p-wave scattering, whose strength can be tuned
using a p-wave Feshbach resonance \cite{EZhaoPRL08}. It is argued that $U$ can be increased to the
order of the recoil energy $E_R$ in the present experiment \cite{CWuPRB08}.

\begin{figure}
\includegraphics[width=0.88\columnwidth]{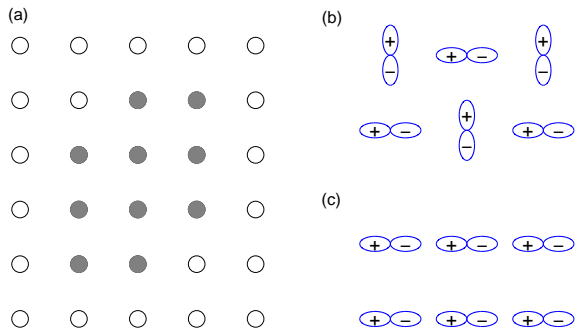}
\caption{(Color online) 
  Reference cluster for the VCA calculation (a) consisting of $L=10$
  sites (gray).  Schematic representation of ``antiferromagnetic''
  (b), and of ``ferromagnetic'' orbital orders (c).
  }
\label{fig1}
\end{figure}

The Variational Cluster Approach (VCA) \cite{po.ai.03,da.ai.04} is an
extension of Cluster Perturbation Theory (CPT)
\cite{gr.va.93,se.pe.00,ov.sa.89}.  Within CPT, the original lattice is
divided into a set of disconnected clusters, and the intercluster hopping
parameters are treated perturbatively.  Within VCA, additional (``virtual'')
single-particle terms are added to the cluster Hamiltonian, to obtain a
so-called reference system, and then subtracted perturbatively.  (So that if
the perturbative treatment was exact, results would not depend on
these terms).
These single-particle terms can contain ``Weiss'' fields to describe a
particular ordered state, but also other Hamiltonian parameters, such as, for
example, an offset in the chemical potential between the cluster and
the lattice.  
The ``optimal'' value for
these variational parameters is determined, in the framework of Self-energy
Functional Approach (SFA) \cite{pott.03,pott.03.se}, by requiring that the
SFA grand-canonical potential
\begin{eqnarray}
\label{omega}
\Omega=\Omega'+\mathrm{Tr}\ln(\mathbf{G}_0^{-1}-\mathbf{\Sigma})^{-1}-\mathrm{Tr}\ln\mathbf{G}'
\end{eqnarray}
is stationary within this set of variational parameters.  Here,
$\mathbf{G}_0$ is the non-interacting Green's function, $\Omega'$,
$\mathbf{\Sigma}$, and $\mathbf{G}'$ are the grand-canonical potential, self
energy, and Green's function of the reference system, respectively.  In this paper, a
$L=10$ sites cluster (Fig.~\ref{fig1}a) is chosen as a reference system,
and
is solved exactly by Lanczos diagonalisation method to obtain the
reference self-energy.
All our calculations are performed at zero temperature for
the well-known difficulty of including the temperature effect into Lanczos method.
Since we are looking for orbital ordering, a 
orbital ferromagnetic or
antiferromagnetic 
field is used
as a variational parameter, in addition to the cluster on-site energy.  The
latter is necessary in order to obtain a thermodynamically consistent
particle density~\cite{ai.ar.05,ai.ar.06}.

The trace in \eqqref{omega} implicitly contains a sum over Matsubara
frequencies which needs to be carried out with high accuracy.
In connection with a Lanczos diagonalisation of the cluster
Hamiltonian this can be done by means of 
the sum over the single-particle excitation energies obtained by the
{\em band Lanczos}~\cite{bandlanczos} method, as explained in 
Ref.~\onlinecite{ai.ar.06.vc} (see also Ref.~\cite{za.ed.02}).
Alternatively, the same accuracy can be obtained more efficiently by an 
 integration over an appropriate
contour of the complex frequency plane, as discussed in
Appendix~\ref{freq}. 
Notice that although
the contour (see Fig.~\ref{contour}) mainly runs at a finite distance
$\delta$ from the real axis in order to avoid sharp structures in the spectral
function in the $\delta\to0$ limit, the procedure is exact. 
There is no need to carry out a $\delta \to 0$ extrapolation:
this is exactly contained in 
the additional contributions from the ``vertical''
paths ( \cc3, \ccp3, \cc5, \ccp5, \cc7, \ccp7 in Fig.~\ref{contour})
(see App.~\ref{freq} for details).

\begin{figure}
\includegraphics[width=0.76\columnwidth]{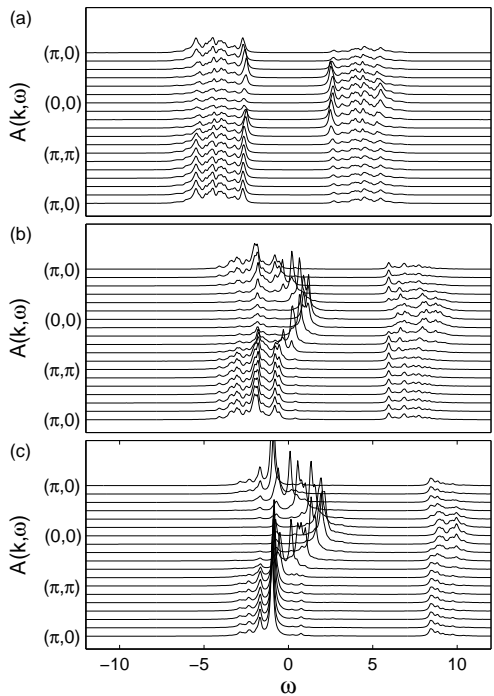}
\caption{Single-particle spectral function $A(\mathbf{k},\omega)$ of $p_x$
  orbital at various fillings
 From  (a) to (c) the fillings of
  the system are 1.0, 0.8, and 0.6, respectively. The interaction
  is $U=8t_{\parallel}$ and the transverse hopping is
  $t_{\perp}=-0.05t_{\parallel}$.}
\label{fig2}
\end{figure}

\section{Results}
\label{results}

\subsection{Filling dependent single-particle spectral function}

In order to gain insight on the physical properties of the
spinless p-band Hubbard model (Eq.~(\ref{spbh})), 
in this section, we calculate its single-particle spectral function
using VCA.
The VCA has been shown to be an effective method to
evaluate the single-particle~\cite{da.ai.04,se.la.05} and
two-particle~\cite{br.ar.08u}
 spectral function of Hubbard-like models.
 The
filling-dependent spectral function of the $p_x$ orbital with interaction
$U=8t_{\parallel}$ is displayed in Fig.~\ref{fig2}. 
By symmetry reasons, the spectrum of the $p_y$ orbital in the
non-ferromagnetic phase is obtained by
 interchanging  $k_x$ with $k_y$.
The spectra at different fillings are obtained in the respectively
stable phase, according to the phase diagram displayed in
Fig.~\ref{fig4} (see Sec.~\ref{phase}).
 The spectrum of the $p_x$ orbital 
is almost $k$-independent in the $k_y$ direction of the Brillioun Zone (BZ), since the dispersion
is nearly 1D. This is quite obviously due to the fact that the
transverse hopping ($t_{\perp}=-0.05t_{\parallel}$) is very small. For all
fillings, we can clearly recognize the upper and lower Hubbard bands with a
gap of the order of $U$.

\begin{figure}
\includegraphics[width=0.76\columnwidth]{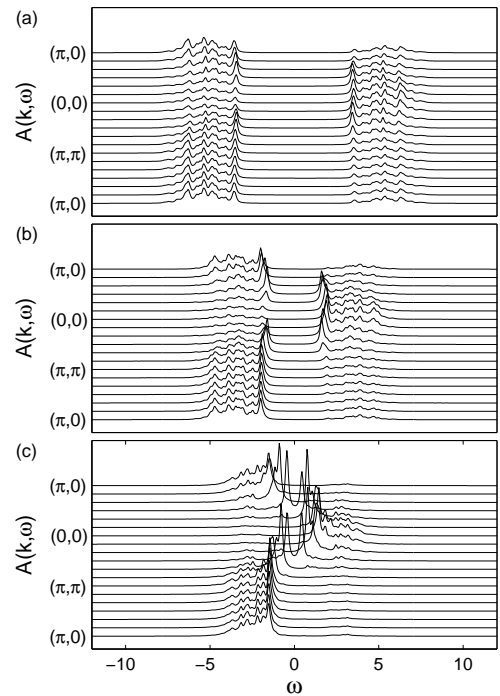}
\caption{Single-particle spectral function $A(\mathbf{k},\omega)$ of
  the $p_x$
  orbital at half-filling and for different values of the interaction
  $U$. 
Specifically, we have 
$U=10t_\parallel$ (a), $6t_\parallel$ (b),
  and $3t_\parallel$ (c).}
\label{fig3}
\end{figure}

At half-filling, the spectrum has a ladder structure (see
Fig.~\ref{fig2}a), which is also characteristic of the  $t-J^z$ model
\cite{EZhaoPRL08,da.wo.08,ma.ho.91}. However, the spectrum is slightly
dispersive in the $k_x$ direction of the BZ, that is, a hole or particle
is not localized but moves coherently through the lattice. The small
dispersion can be explained by including a three-sites term in the
$t-J^z$ Hamiltonian \cite{da.wo.08}. More spectra at half-filling and
for different interactions are given in Fig. \ref{fig3}. The gap
between upper and lower Hubbard bands decreases as the interaction
decreases. At the same time, the bandwidth become larger because a hole
(or particle) is easier to move when the interaction is
smaller.

Away from half-filling, the quasi-particle spectrum becomes strongly
dispersive (see Fig.~\ref{fig2}). The shape of the spectrum is similar
to that of 1D free particles, but with a strongly renormalized
bandwidth. The bandwidth becomes larger and larger when going away from
half-filling, which means that particles can move easier.
Another feature that can be seen in Fig.~\ref{fig2} is the spectral weight transfer
phenomenon between the upper and lower Hubbard bands, which is also
observed in the usual single-band Hubbard model~\cite{es.me.91}. 
For fillings below
half-filling (Fig.~\ref{fig2}b,c), the spectrum is transferred
to the lower Hubbard band. This is because at low density
the particles have less chance to doubly occupy the same site and
therefore have a smaller probability to be in the upper Hubbard
band~\cite{es.me.91}. 
Of course, for filling  above
half-filling the
situation is reversed due to particle-hole symmetry.

\subsection{Antiferromagnetic orbital order}
\label{phase}

\begin{figure}
\includegraphics[width=0.98\columnwidth]{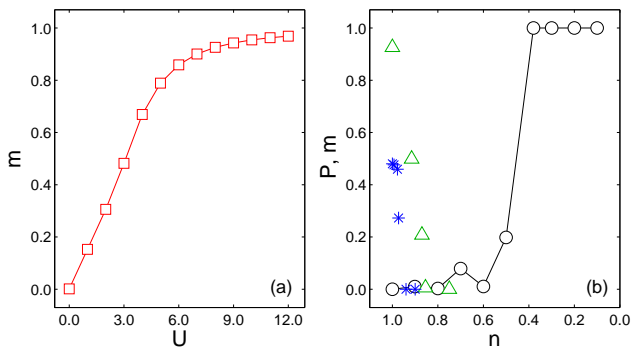}
\caption{
(Color online)
(a), Staggered orbital order parameter $m$ ($\square$) as a function of
  interaction $U$ at half-filling. 
(b), Orbital polarization $P$ at $U=8t_\parallel$ ($\circ$)
   and staggered orbital order parameter $m$ at $U=8t_\parallel$
   ($\triangle$) and  $U=3t_\parallel$ ($\ast$) as a function of filling
   $n$.
}
\label{fig4}
\end{figure}

In this section, we discuss the ``antiferromagnetic'' orbital 
order in this model as a
function of filling.
At half-filling
and in the strong-coupling limit $U\gg t_{\parallel}$, the Hamiltonian Eq.
(\ref{spbh}) can be reduced to a superexchange $t-J^z$ model with a positive
exchange energy $J=2t_{\parallel}^2/U$ \cite{EZhaoPRL08,CWuPRL08,da.wo.08}.
Therefore, the Mott state favors a staggered  (``antiferromagnetic'') 
orbital order (see
Fig.~\ref{fig1}b). To study the orbital order within VCA, we add a
``virtual'' staggered orbital field,
$H'_{AF}=h'_{AF}\sum_\mathbf{r}(n_{\mathbf{r},x}-n_{\mathbf{r},y})
e^{i\mathbf{Q}\cdot\mathbf{r}}$ with
$\mathbf{Q}=(\pi,\pi)$, to the cluster Hamiltonian $H'$. 
As explained in Sec.~\ref{vca}, this term is then subtracted
perturbatively, and the coefficient is determined by optimizing the
grand-canonical potential \eqqref{omega}.
 The corresponding staggered
orbital order parameter, 
$m=\sum_\mathbf{r}(\langle n_{\mathbf{r},x}\rangle-\langle
n_{\mathbf{r},y}\rangle)e^{i\mathbf{Q}\cdot\mathbf{r}}$, 
is then calculated and
plotted in Fig.~\ref{fig4}a as a function of the interaction $U$. One can see
that the order parameter $m$, which is non-zero for any 
finite $U$, is increasing as $U$
increases and  approaches unity in the strong coupling limit. This result
supports the existence of the ``antiferromagnetic'' orbital order at
half-filling.

The ``antiferromagnetic'' orbital order is destroyed by doping the system
away from
half-filling. 
This is 
illustrated in Fig. \ref{fig4}b, where the staggered
orbital order parameter $m$
is plotted as a function of
filling $n$ at different interactions
$U=8t_\parallel$ (denoted by $\triangle$) and $U=3t_\parallel$ (denoted by $\ast$).
Fig. \ref{fig4}b shows that the order parameter $m$ decreases sharply
when $n$ decreases, and disappears completely ($m=0$) at fillings $n \approx
0.85$
and $n \approx 0.94$ for $U=8t_\parallel$ and $U=3t_\parallel$, respectively.
We conclude that
the ``antiferromagnetic'' orbital order at large $U$ is more difficult to destroy
than at small $U$. 
After $m=0$, the system enters a featureless ``paramagnetic'' orbital state.

\subsection{Orbital order at low filling}

As shown in the spectrum (see Fig. \ref{fig2}, \ref{fig3}), 
the spinless p-band model has strong 1D character
for each orbital due to the anisotropic hopping, and therefore its band structure
has a Van Hove singularity near the band edge \cite{LWangPRA08}.
It is interesting to see whether
or not this singularity can produce ``ferromagnetic'' orbital 
order at low filling (shown schematically in
Fig. \ref{fig1}c) \cite{Fazekas}.

The question is subtle because the same 
Van Hove singularity present
in the 1D single-band Hubbard model at low filling 
is not sufficient to obtain ferromagnetism.
In particular, an Hartree-Fock argument provides the wrong conclusion
that the ferromagnetic state should be lower in energy than the
paramagnetic state at sufficiently low
densities and large $U$
in one dimension.
This results is indeed contradicted by the rigorous Lieb-Mattis
theorem \cite{HTasakiPTP98}, which excludes ferromagnetism for the 1D
Hubbard model, 
as well as
by an accurate analysis based on the Gutzwiller wave function.

While the  Lieb-Mattis
theorem does not apply to the present 
p-band
model, we investigate
here whether or not 
an instability of the totally polarized ferromagnetic
state towards a variational, less polarized, and, ultimately,
paramagnmetic wave function can be found for the 
p-band
model in the
low-density and $U\to\infty$ limit.
As trial wave functions for the less polarized state we use the
Gutzwiller wave function, as well as a more general one, i. e. with
lower energy. Despite this, we find that the totally polarized
state, which, of course, can be solved exactly for an onsite
interaction, always has the lowest energy.
While we were not able so far to prove that the totally polarized
state is the most stable one at sufficiently low 
filling,
the fact that
we have used a quite general trial wave function
makes us confident that there should be no 
wave functions with a lower
energy than the totally polarised state.

We consider a 
p-band
model with $N$ particles in a finite $L \times L$ square
lattice with periodic boundary conditions 
(PBC).
For simplicity, we take $t_\perp=0$ and 
$U=\infty$.
 Quite generally, we can
expect that if the ferromagnetic phase has a lower energy with a
finite gap to the paramagnetic state
for these values of 
$t_\perp$ 
and $U$,
 its
stability region 
should extend to some finite 
$t_\perp$ and $U$.

If $N \le L$, 
it is quite clear that the lowest energy is obtained by putting all
particles in the same orbital (say, $p_x$) on different ``rows''. In
that case, each particle moves independently on its row, so that the
kinetice energy is minimal and the interaction energy is zero.
 However, this 
cannot lead to the conclusion that the ferromagnetic state is stable 
at sufficiently low but finite density, since for $N \le L$ the
density vanishes in the thermodynamic limit.
The crucial question is what happens for $N=L+1$, 
i.e. is it more convenient energetically to put the next particle in
one of the already occupied rows in the $p_x$ orbital, or to put it in
a ``column'' in the $p_y$ orbital?
If the particle is added to the $p_x$
orbital, the system is still in a full ferromagnetic state, 
and the energy change
$\Delta E_1$ of this state
with respect to the ground state with $N=L$,
$|L\rangle_x$,
 (which has 
energy $E_0=-2t_{\parallel}L$), is given by
\beq
\label{de1}
\Delta E_1=
-2t_{\parallel}\cos(\frac{2\pi}{L})
\approx
-2t_{\parallel}(1-\frac{2\pi^2}{L^2}) \;,
\eeq
where, in the last term, we have taken the large-$L$ limit.
If we add the particle to the $p_y$ orbital on one of the columns (no
matter which one), 
the lowest-energy state cannot be determined exactly. Therefore, we 
approximate it by
a trial wave function. 
The simpliest one is the Gutzwiller wave
function
\beq
\label{gwf}
|\psi \rangle =\prod_\mathbf{r} (1-n_{\mathbf{r},x}n_{\mathbf{r},y})
   d^\dag_{(x=0,q_y=0),y}|L\rangle_x
\eeq
where 
$d^\dag_{(x,q_y),y}$
 creates a particle on $p_y$ orbitals
on ``column'' ($x$) with $y$ wave vector 
 vector $q_y$.
The energy increase 
can be easily evaluated as
\beq
\Delta E_2=-2t_{\parallel}(1-\frac{2}{L-1})
\eeq
Clearly, this energy is larger than \eqref{de1}. The reason for this
is that
for the 
row where the $p_y$ particle sits, the Gutzwiller wave function of the
$p_x$
particle has a sharp jump at
the position of the $p_y$ particle, see Eq. (\ref{gwf}). This leads to an 
increase of $2t_{\parallel}/(L-1)$ in the kinetic energy of the $p_x$ particle. 
A natural improvement 
consists in replacing
the wave function of this row by a smooth
sine function, $\sin(\pi x/L)$, which has much smaller energy
increase. However, 
the overlap of this sin function with the wave
function of the other rows, which is $\sqrt{1/L}$, is small, resulting in a 
large kinetic energy of the
$p_y$ particle. A proper choice is to linearly combine these two functions.
This leads to the
 trial wave function: 
\beqn
\label{twf}
|\psi \rangle &=& \frac{1}{L}
 \sum_m  c^\dag_{(0,m),y} \\ \nonumber
         &\times &\prod_n \sum_l 
           \Big[ a_{m-n}+ \sqrt{2}\
                 b_{m-n}\sin(\frac{\pi l}{L}) \Big]
           c^\dag_{(l,n),x}|0\rangle
\eeqn
where 
$(l,m)$ explicitly denotes the 2D coordinates of a position
$\mathbf{r}$, and $l,m,n=0,1,\cdots,L-1$.
Since we are using PBC, our assumption that 
the $p_y$ particle is in
the $0th$ column
does not lead to a
 loss
 of generality.
The coefficients $a_{m-n}$ and $b_{m-n}$
can be 
chosen, 
  for simplicity, 
to be $a_{m-n}=\sin^2\frac{(n-m)\pi}{L}$, and
$b_{m-n}=\cos^2 \frac{(n-m)\pi}{L}$.
This choice does not affect our conclusions, as discussed below.
The
energy 
increase for the state in \eqref{twf} 
is given by:
\beq
\label{de3}
\Delta E_3 = -2t_{\parallel}(1-\frac{\alpha}{L} + 
O(L^{-2})) \;,
\eeq
where $\alpha \approx 0.5$ is a constant.
A comparison of
 the energies of 
this trial wave function with the 
fully polarized ferromagnetic state, 
whose energy increase is $-2t_{\parallel}+O(L^{-2})$, 
shows that the latter has a lower energy.
Notice that more general forms of the coefficients
$a_{m-n}$ and $b_{m-n}$ 
do not change this conclusion, as they
merely modify the coefficient $\alpha$
 in
 Eq.~(\ref{de3}),
 which, however, remains nonzero and
positive.

The above results show that,
althought the spinless p-band model has strong 1D character, it is different
from the 1D Hubbard model. 
This can be seen by constructing 
a trial wave function 
in a similar way to
Eq.~(\ref{twf})
for the 1D Hubbard model with two particles \cite{gebh.97}.
In this case, the total energy
for PBC is $-4t_{\parallel}+C/L^2+O(L^{-4})$ 
with $C=2\pi^2 t_{\parallel}$ for
the paramagnetic state and 
$C=4\pi^2 t_{\parallel}$ for the ferromagnetic state \cite{gebh.97},
i.e., the ferromagnetic state is unstable.
This situation is quite different from 
the partly polarized state of the p-band model 
presented above with one particle
in the $p_y$ orbital and $L$ particles in $p_x$ orbital. In this case,
the motion of the
$p_y$ particle is hindered by,  and, at the same
time, affects the motion of the other $L$ particles in the $p_x$
orbitals.
This 
leads to a much larger energy
increase than in the fully polarized ferromagnetic state.

After having discussed the stability of the ferromagnetic phase from a
more accurate point of view, we return to the results of the VCA
approximation in the
low-filling region.  
In the ferromagnetic case
it is necessary to introduce a different
on-site energy between the two orbitals as a variational parameter.
This is equivalent to using the cluster chemical potential and a
``ferromagnetic'' field.
 In the fully polarized case, the saddle point
is given by the on-site energy of the empty orbital approaching infinity.
To describe 
the ferromagnetic phase, we evaluate the orbital
polarization $P\equiv(n_x-n_y)/(n_x+n_y)$, where $n_x$ and $n_y$ are
the average occupations of the $p_x$ and $p_y$ orbitals. 
Results for $P$ as a function of filling for $U=8 t_\parallel$ are plotted in
Fig. \ref{fig4}b (denoted by $\circ$).
The orbital polarization $P$ is calculated at the
respectively stationary point of $\Omega$ in each phase. 
As in Fig. \ref{fig4}b,
$P$  vanishes at half-filling 
and remains essentially zero down to a filling of $n\approx 0.6$.
For $n<0.6$,
 $P$  rapidly increases as
$n$ decreases, and 
rapidly
saturates ($P=1$) at $n\approx 0.38$
indicating a full ferromagnetic orbital order state.

We should stress that one
must be careful when interpreting the VCA results at
 low filling. 
First,  we cannot exclude that finite-size effects, originating from
the limited size of the  reference cluster,
could affect the ordered state found in our calculation.
This could be the case when the exact self-energy is long ranged, so
that it cannot be accurately described
 by the self-energy of a small reference system.
Second, the density obtained by
VCA shows small discontinuities 
when the reference cluster changes its particle
number~\cite{ba.ha.08}. 
Therefore, it is difficult for VCA to determine the
exact critical point for the onset of ferromagnetic orbital
ordering as a function of filling.

\begin{figure}
\includegraphics[width=0.96\columnwidth]{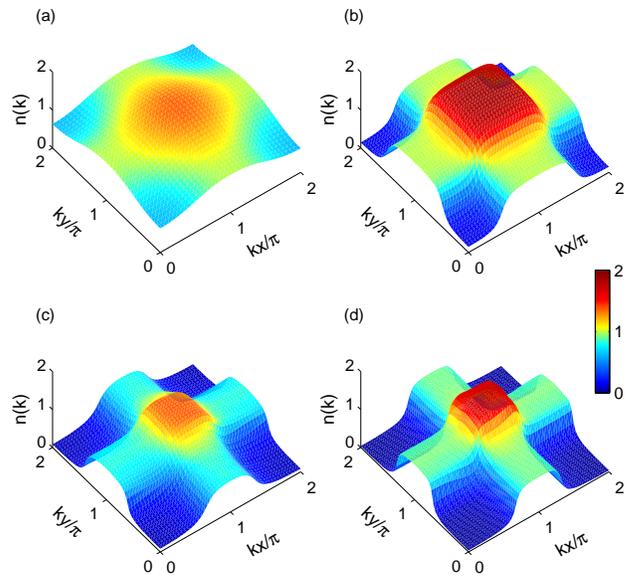}
\caption{(Color online)
Momentum distribution $n(k)$ for different values of the interaction $U$ and
of the  filling $n$, with $t_{\perp}=-0.05t_{\parallel}$. (a), $U=10t_\parallel$,
  $n=1.0$; (b), $U=3t_\parallel$, $n=1.0$; (c), $U=10t_\parallel$, $n=0.6$;
  (d), $U=3t_\parallel$, $n=0.6$.
   }
\label{fig5}
\end{figure}

Summarizing this section, 
our combined VCA and variational results are a strong
indication, although not a proof, for the presence of an orbital ferromagnetic state at
low-density and sufficiently large $U$ in the p-band model. 
 An exact proof for the absence or
existence of ferromagnetism at low densities in the p-band model
(similarly to the Hubbard model \cite{Fazekas}) 
would be welcome. However, it 
is beyond  the goal of
the present paper.

\subsection{Momentum distribution}

In this section, we present results for the momentum distribution,
as this quantity 
 is directly accessible experimentally~\cite{Kohl05}, and can be
used to detect the possible occurrence of orbital ordering.
Our results
are shown in Fig. \ref{fig5}. For
half-filling and small values of the interaction $U$ (Fig. \ref{fig5}b), the momentum
distribution can approximately be seen as the 
superposition of two 1D noninteracting gases traveling in
the two directions $x$ and $y$. 
Double occupations are present in $k$ space in the middle
square region of the Brillioun Zone. 
While double occupation is allowed for small $U$, 
it is strongly suppressed for strong interactions.
Therefore, for large $U$,
 the momentum distribution is
flattened and covers the whole BZ (see Fig. \ref{fig5}a). 
By decreasing the filling away from half filling
the suppression of double occupation weakens, as can be seen in 
Figs. \ref{fig5}c, and d for $n=0.6$.
 The
reason is that the system has a strongly dispersive spectrum (see
Fig. \ref{fig2}c) and, therefore,  it is no longer in a Mott state. 
The fact that particles can
move quite free in the lattice, gives rise to the possibility of double
occupation even at large interactions.

Finally, we briefly discuss the experimental signatures. The momentum
distribution of fermions in the excited p-band (see Fig. \ref{fig5}) is
different from that of fermions in the lowest s-band \cite{Kohl05}, e.g., in the weak
interacting regime. This can be directly observed in the time of flight (TOF)
images. The antiferromagnetic orbital order can be detected by analyzing
the noise correlation function from TOF images. In the noise correlation
spectrum, the s-band fermions
produce the antibunching dips at the usual
reciprocal wave vector of square lattice \cite{Altman04,Rom06}. However, the
p-band fermions in the antiferromagnetic orbital order state contribute new dips
at the reciprocal wave vector of doubled unit cell \cite{EZhaoPRL08}.

\section{Conclusion}
\label{conc}

In summary, we have studied a model for spinless p-band fermions in optical lattices
using the Variational Cluster Approach, 
and, partly, a variational
wave function.
We have computed its
single-particle spectral function in a wide range of fillings and found a
strongly dispersive spectrum at incommensurate fillings. 
By calculating the
staggered orbital order parameter, we showed that the
system is in a staggered (``antiferromagnetic'') orbital state at
half-filling,
which is destroyed by doping and evolves into a paramagnetic state.
In the low-density limit and for $U=\infty$, we studied
 the stability of a fully-polarized 
ferromagnetic state by constructing a trial wave
function, which extends the Gutzwiller trial state.
In contrast to the one- and two-dimensional Hubbard model
we did not find an instability of the ferromagnetic state towards the
paramagnetic solution. 
In particular, for the trial wave function
of Eq.~(\ref{twf}) (which is more general than the Gutzwiller wave function), 
the ferromagnetic
state is lower in energy than the paramagnetic one.
Finally, we have computed, by VCA, the
momentum distribution and studied its evolution as a function
of interaction and filling.

\begin{acknowledgments}
We thank M. Daghofer for helpful discussions, as well as
 H. Allmaier for precious assistance in the VCA code.
This work was partly supported by the Austrian Science Fund
(FWF P18551-N16).
\end{acknowledgments}

\appendix

\section{Frequency integration}
\label{freq}

The sum over Matsubara frequencies in \eqqref{omega} can be carried out
either (i) directly (see also Ref.~\cite{sene.08u}) 
or (ii) by the usual procedure \cite{book} of
distorting the contour to the real axis. For a numerical sum, and
especially for the corresponding integral at $T=0$, both procedures
present their
advantages and disadvantages. In case (i) one should first 
extract the asymptotic (for large $i \omn$) part of the integrand
and carry out the corresponding sum/integral analytically.
In case (ii) there is no such problem, as
 the contribution to the integral at an infinitesimal
distance $\delta\to 0$ from  the real axis
is nonzero only within the region where the spectral function is
nonzero. However, due to the pole structure of the integrand,
 one has to take a finite $\delta$ for
numerical purposes. This reduces the precision and introduces
additional complications coming from the fact that at large \om\ the
integrand goes like $1/\om^2$.

\begin{figure}[h]
\includegraphics[width=0.99\columnwidth]{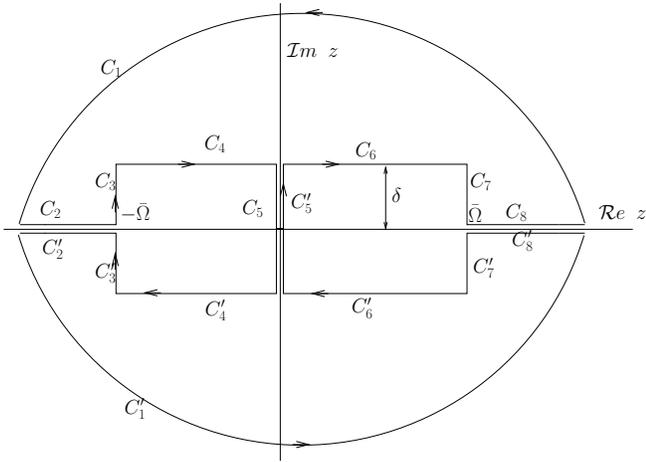}
\caption{\label{contour}
Contour in the complex plane in which the frequency integration is carried out.
}
\end{figure}

The best solution is to distort the integral to the contour
 indicated in  
Fig.\ref{contour}. 
For a sum over Matsubara frequencies $\omn = 2 \pi T (n +
\frac12)$ of a function $g(z)$ of the complex variable $z$, which is
analytic everywhere except on the real axis , we have
\beq
\label{sumn}
T \sum_{n=-\infty}^{+\infty} e^{i \omn 0^+} g(i \omn) =
- \frac{1}{2 \pi i} \oint_{C} e^{z  0^+} f_F(z) g(z) d z \;.
\eeq
Here, $C$ is the usual contour of the complex plane encircling the
Matsubara frequencies $i \omn$,
$f_F(z) = (\exp \frac{z}{T} +1 )^{-1}$ is the Fermi function,
and $0+$ is a positive infinitesimal.
 With the usual conditions that
$g(z)\to 0$
 for $|z| \to \infty$, and that 
\beq
\label{gzcc}
g(z^*)=g(z)^* \;,
\eeq
 we can further
 distort the contour $C$ to the contour indicated in Fig.~\ref{contour}.
Here,  \cc1 and \ccp1 are semicircles at infinity, so that their
contribution vanish, \cc2, \cc8, and  \ccp2, \ccp8 are infinitesimally
close to the real axis, as in the usual procedure. 
\cc5 and \ccp5 are infinitesimally close to the imaginary axis, while
\cc4, \cc6, and \ccp4, \ccp6 can have an arbitrary finite distance $\DD$
from the real axis.
We take $\bar \Omega$ as some upper limit of the spectrum, i.\,e.,
\beq
\label{upperlimit}
\im g(\om + i 0^+) = 0 \quad \hbox{ for any } \quad |\om| > \bar\Omega \;.
\eeq
By calling
\beq
F(z) \equiv g(z) f_F(z)
\eeq
we can first write the contributions to \eqqref{sumn}
from the ``horizontal'' paths:
\beq
S_h=
- \frac{1}{\pi} \int_{\cc2+\cc4+\cc6+\cc8} \im F(z) \ d z \;.
\eeq
Notice that for $\DD\to 0$ one recovers the usual expression
\cite{book}, and there is obviously no contribution from the
``vertical'' paths.
The contributions from $\cc2$ and $\cc8$ vanish, due to \eqqref{upperlimit}
($f_F(z)$ is analytic across the real axis).
Therefore, we are left with
\beq
\label{sh}
S_h=
 - \frac{1}{\pi} \int_{-\bar\Omega}^{\bar\Omega} \im\left[g(\om+i \DD)
  f_F(\om+i \DD)\right] 
 d \om \;.
\eeq
The advantage of taking a finite $\DD$ is that the integrand is smooth
and one only needs few \om\ points in the numerical integration 
in order to achieve a good accuracy, in contrast
to the conventional case of small $\DD$. 
For $T=0$ \eqqref{sh} reduces to
\beq
S_h = - \frac{1}{\pi} \int_{-\bar\Omega}^{0} \im g(\om+i \DD)
 d \om \;.
\eeq
The rest of the integral is given by the ``vertical'' paths, for
example the contribution from \cc3, \ccp3 is given by
\beqn
\label{c3}
&&
-\frac{1}{2 \pi i} \int_{\cc3+\ccp3} \cdots =
\\ \nonumber &&
\frac{i}{2 \pi} \left[ \int_{0}^{\DD} F(-\bar\Omega+i x) \ i \ d  x
 -
\int_{0}^{\DD} F(-\bar\Omega-i x) \ (-i) \ d  x \right]=
\\ \nonumber &&
-\frac{1}{\pi} \int_{0}^{\DD} \re F(-\bar\Omega+i x) d x
\eeqn
and similarly for the contribution from \cc7, \ccp7:
\beq
\label{c7}
-\frac{1}{2 \pi i} \int_{\cc7+\ccp7} \cdots = 
\frac{1}{\pi} \int_{0}^{\DD} \re F(\bar\Omega+i x) d x \;.
\eeq
The latter contributions vanishes for $T=0$ or can be made
exponentially small by taking $\bar\Omega/T \gg 1$.
The contribution from the ``central'' vertical paths
\cc5,\ccp5 is simply given by the original sum over Matsubara
frequencies, however only for $|\omn| < \DD$  (we must be wise and 
choose $\DD$ not to coincide with a Matsubara frequency for $T>0$).
Denoting by $\omega_{n_{max}}$ the corresponding maximum frequency, we have
\beq
\label{c5}
-\frac{1}{2 \pi i} \int_{\cc5+\ccp5} \cdots = 
2 T \sum_{n=0}^{n_{max}} \re g(i \omn)
\eeq
which for $T=0$ becomes
\beq
\label{c5t0}
\frac{1}{\pi} \int_{0}^{\DD} \re g(i x ) d x \;.
\eeq
The contributions \eqqref{c3}, \eqqref{c7}, \eqqref{c5} are the
additional integrals to be carried out to compensate for the
nonvanishing value of $\DD$. We stress that the result is exact for
any (even large) value of $\DD>0$. The numerical advantage is that
the integrand is everywhere smooth except
for small temperatures and on the path \cc5 near $\omn=0$ whenever
$g(\om)$ has poles close to $\om=0$.
Moreover, all integrals are carried out in a finite domain, so there
is no need to carry out extrapolations.


\end{document}